\documentclass[aps,prb,showpacs,amsmath,twocolumn,amssymb,superscriptaddress,letterpaper]{revtex4}
\usepackage{times}
\usepackage{amsfonts}
\usepackage{mathrsfs}
\usepackage{graphicx}
\usepackage{dcolumn}
\usepackage{bm}
\usepackage{color}

\usepackage[colorlinks,bookmarks=false,citecolor=blue,linkcolor=red,urlcolor=blue]{hyperref}
\def\be{\begin{equation}}       \def\ee{\end{equation}}
\def\bea{\begin{eqnarray}}      \def\eea{\end{eqnarray}}

\begin{document}
\title{The effect of As-Chain layers in  CaFeAs$_2$}

\author{Xianxin Wu}
\affiliation{ Institute of Physics, Chinese Academy of Sciences,
Beijing 100190, China}

\author{Congcong Le}
\affiliation{ Institute of Physics, Chinese Academy of Sciences,
Beijing 100190, China}

\author{Yi Liang}
\affiliation{ Institute of Physics, Chinese Academy of Sciences,
Beijing 100190, China}

\author{Shengshan Qin}
\affiliation{ Institute of Physics, Chinese Academy of Sciences,
Beijing 100190, China}

\author{Heng Fan }  \email{hfan@iphy.ac.cn }  \affiliation{ Institute of Physics, Chinese Academy of Sciences,
Beijing 100190, China}

\author{Jiangping Hu  } \email{jphu@iphy.ac.cn}  \affiliation{
Institute of Physics, Chinese Academy of Sciences, Beijing 100190,
China}\affiliation{Department of Physics, Purdue University, West
Lafayette, Indiana 47907, USA}

\date{\today}

\begin{abstract}
The new discovered iron-based superconductors have  chain-like As layers.  These layers generate an additional 3-dimensional hole pocket and cone-like electron pockets. The former is attributed to the Ca $d$ and As1 $p_z$ orbitals and the latter are attributed to the anisotropic Dirac cone, contributed by As1 $p_x$ and $p_y$ orbitals.  We find that large gaps on these pockets open in the collinear antiferromagnetic ground state of  CaFeAs$_2$, suggesting that  the chain-like As layers are strongly coupled to FeAs layers.  Moreover due to the low symmetry crystal induced by the  As layers, the bands attributed to FeAs layers  in $k_y=\pi$ plane are two-fold degenerate but in $k_x=\pi$ plane are lifted. This degeneracy is protected by a hidden symmetry $\hat{\Upsilon}=\hat{T}\hat{R}_y$. Ignoring the electron cones,  the materials can be well described  by a six-band model, including five Fe $d$ and As1 $p_z$ orbitals.  We suggest that these new features  may help us to identify the sign change and pairing symmetry in iron based superconductors.

\end{abstract}


\maketitle

\section{introduction}
The discovery of high-temperature superconductivity in
LaFeAsO$_{1-x}$F$_x$\cite{Kamihara2008} has generated considerable
interest in the study of iron pnictides. Many different families of materials have been discovered, including  the 1111 family with T$_c$  exceeding 55 K\cite{Ren2008,WangCao2008}, the 122 family such as (K,Ba)Fe$_2$As$_2$ with T$_c$ around 38 K\cite{Rotter2008,Rotter2008L}, the 111 family such as LiFeAs  with T$_c$  around 20 K \cite{Wang2008,Chu2009,Sasmal2009} and so on. All the iron pnictides discovered so far consist of FeAs layers and various kinds of blocking layers.  Among all these materials,  their Fermi surfaces are qualitatively similar and the lattice symmetry groups are either   $P4/nmm$ or $I4/mmm$ which result in  double  degenerate band structure on the Brillouin zone boundary.  The blocking layers in these materials are mostly insulating so that the electron states from them have little  contribution to electronic physics.

Very recently, two novel iron pnictide materials (Ca,Pr)FeAs$_2$\cite{Yakita2013} and Ca$_{1-x}$La$_x$FeAs$_2$\cite{Katayama2013} have been synthesized and the superconductivity transition temperature are 20 K and 45 K respectively. In angle-resolved photoemission measurements, Liu \emph{et al}. observed three hole-like bands around the Gamma point and one electron-like Fermi surface near the M point\cite{Liu2013}. In contrast to the previous discovered iron based superconductors, Ca$_{1-x}$La$_x$FeAs$_2$ belongs to the monoclinic space group P2$_1$, containing only two operations: the two-fold screw axis along $y$ axis and the identity. It contains alternately stacked FeAs and arsenic layers. The As atoms in arsenic layers form zigzag chains and they are quite different from the anions square network in LaOFeAs. More interestingly, it was predicted that the blocking layers Ba$Pn$ in a hypothetical compound BaFePn$_2$($Pn$=As, Sb) with a similar  structure were predicted to be metallic through first principle calculation\cite{Shim2009}.

In this paper, we perform density functional(DFT) calculation to
investigate the electronic and magnetic structures of CaFeAs$_2$. We
find that CaFeAs$_2$ is a bad metal similar to the other iron based
materials. However, we found that  the CaAs blocking layers are  metallic, which can make a key difference on electronic properties from the other  iron based superconductors. In the paramagnetic state, the blocking layers contribute an additional 3-dimensional   hole pocket  at $\Gamma$ point and four electron cones. The additional hole pocket is contributed by the Ca $d$ and As1 $p_z$ orbitals and the electron cones are attributed to the As1 $p_x$ and $p_y$ orbitals. The ground state of CaFeAs$_2$ is found to be collinear antiferromagnetic. The  gaps on the electron cones are opened in the collinear antiferromagnetic  state, suggesting that  the CaAs layers are strongly coupled to FeAs layers.  Moreover due to the low symmetry crystal induced by the  blocking layers, the bands attributed to FeAs layers  in the $k_y=\pi$ plane are two-fold degenerate, but the degeneracy  in the $k_x=\pi$ plane is lifted. This degeneracy is protected by an hidden symmetry $\hat{\Upsilon}=\hat{T}\hat{R}_y$. Ignoring the electron cones, the materials can be well described  by a six-band model, including five Fe $d$ and As1 $p_z$ orbitals.  We suggest that these new features  may help us to sort out proposals related to pairing symmetries in iron-based superconductors.

The paper is organized as following. In Section~\ref{S1}, the DFT band structures and fermi surfaces in paramagnetic and collinear antiferromagnetic phase are presented. Then, in Sec.~\ref{S2}, we obtain the tight binding models for the As layers and  FeAs layers and the anisotropic Dirac cone is shown. In Sec.~\ref{S3}, we discuss the new features in this material, which may be used to test the pairing symmetry of the iron based superconductors. Finally, in
Sec.~\ref{S4}, we give a summary and provide the main conclusions of
our paper.

\begin{figure}[t]
\centerline{\includegraphics[height=7cm]{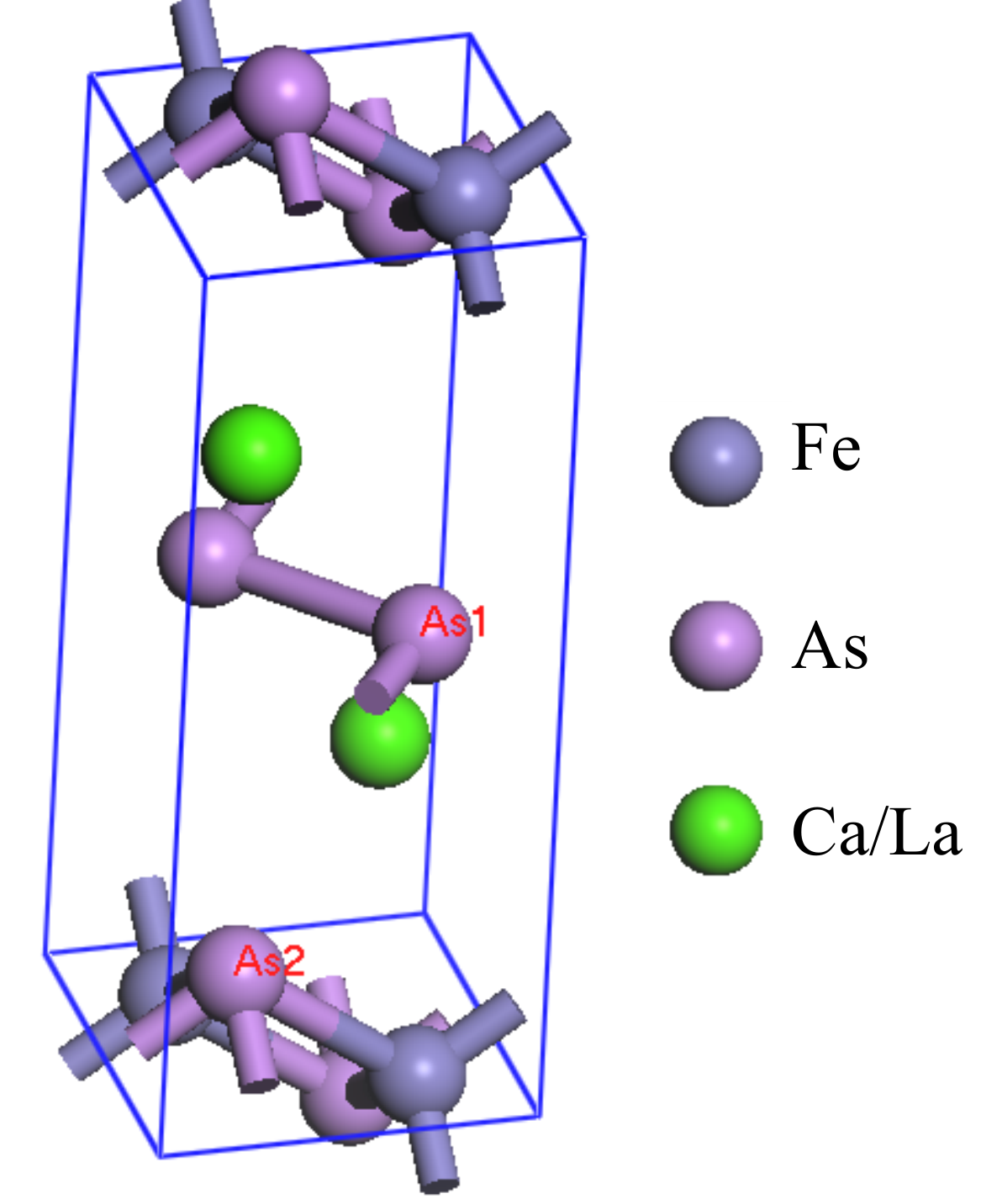}}
\caption{(color online). Schematic view of the structure of Ca$_{1-x}$La$_x$FeAs$_2$.
 \label{model} }
\end{figure}

\section{Electronic structure} \label{S1}
\subsection{Structure and Method}
Our DFT calculations employ the projector augmented wave (PAW)
method encoded in Vienna \emph{ab initio} simulation package(VASP)
\cite{Kresse1993,Kresse1996,Kresse1996B}, and both the local density
approximation (LDA) and generalized-gradient approximation (GGA)\cite{Perdew1996} for
the exchange correlation functional are used. Throughout this work,
the cutoff energy of 500 eV is taken for expanding the wave
functions into plane-wave basis. In the calculation, the Brillouin
zone is sampled in the \textbf{k} space within Monkhorst-Pack
scheme\cite{MonkhorstPack}. The number of these $k$ points are
depending on the lattice: $11\times11\times4$ and $8\times8\times4$ for the paramagnetic phase and collinear antiferromagnetic phase, respectively . 
We relax the lattice constants and internal atomic positions with both LDA and GGA, where the plane wave cutoff energy is 600 eV. Forces
are minimized to less than 0.01 eV/\AA in the relaxation.

The crystal structure of Ca$_{1-x}$La$_x$FeAs$_2$ is shown in Fig.\ref{model}. Although pure CaFeAs$_2$ has not been experimentally synthesized, we perform the calculation for the parent compound CaFeAs$_2$. The optimized and experimental structural parameters of CaFeAs$_2$ are summarized in Table
\ref{structure}. The calculated lattice constants using both LDA and GGA are in good agreement with the experiment data. However, the calculated length of Fe-As2 bond in the paramagnetic phase are smaller than the experimental value. The GGA calculated As2 height differs from the experimental data by 0.11 \AA. This underestimated As2 height has been also
noted in the BaFe$_2$As$_2$\cite{Singh2008} and LaFeAsO
\cite{Yin2008,Mazin2008} compounds and may be related to the strong spin fluctuation, which is beyond DFT calculation. As the calculated lattice parameters using GGA are more closer to the experimental data, we perform the following calculation using the lattice parameters obtained by relaxation with GGA.

\begin{table}[bt]
\caption{\label{structure}%
The optimized and experimental structural parameters of CaFeAs$_2$ using
LDA and GGA in the paramagnetic phase. The space group is P2$_1$(No. 4). The experimental data is for Ca$_{1-x}$La${_x}$FeAs$_2$. Deviations between the optimized and experimental values are given in parentheses in \% .
}
\begin{ruledtabular}
\begin{tabular}{cccc}
& LDA & GGA & EXP \\
 \colrule
$a$(\AA) & 3.851(-2.43) & 3.957(0.26) & 3.947 \\
$b$(\AA) & 3.800(-1.86) & 3.895(0.59) & 3.872 \\
$c$(\AA) & 9.758(-5.46) & 10.075(-2.39) & 10.321 \\
$\beta$($\circ$) & 91.1(-0.35) & 91.3(-0.11) & 91.4 \\
Fe-As2(\AA) & 2.27; 2.30 & 2.32; 2.36 & 2.41; 2.42  \\
As1-As1(\AA) & 2.52; 2.90 & 2.56; 3.01 & 2.53; 3.02 \\
\end{tabular}
\end{ruledtabular}
\end{table}

\subsection{Band and DOS}
\begin{figure}[tb]
\centerline{\includegraphics[height=12 cm]{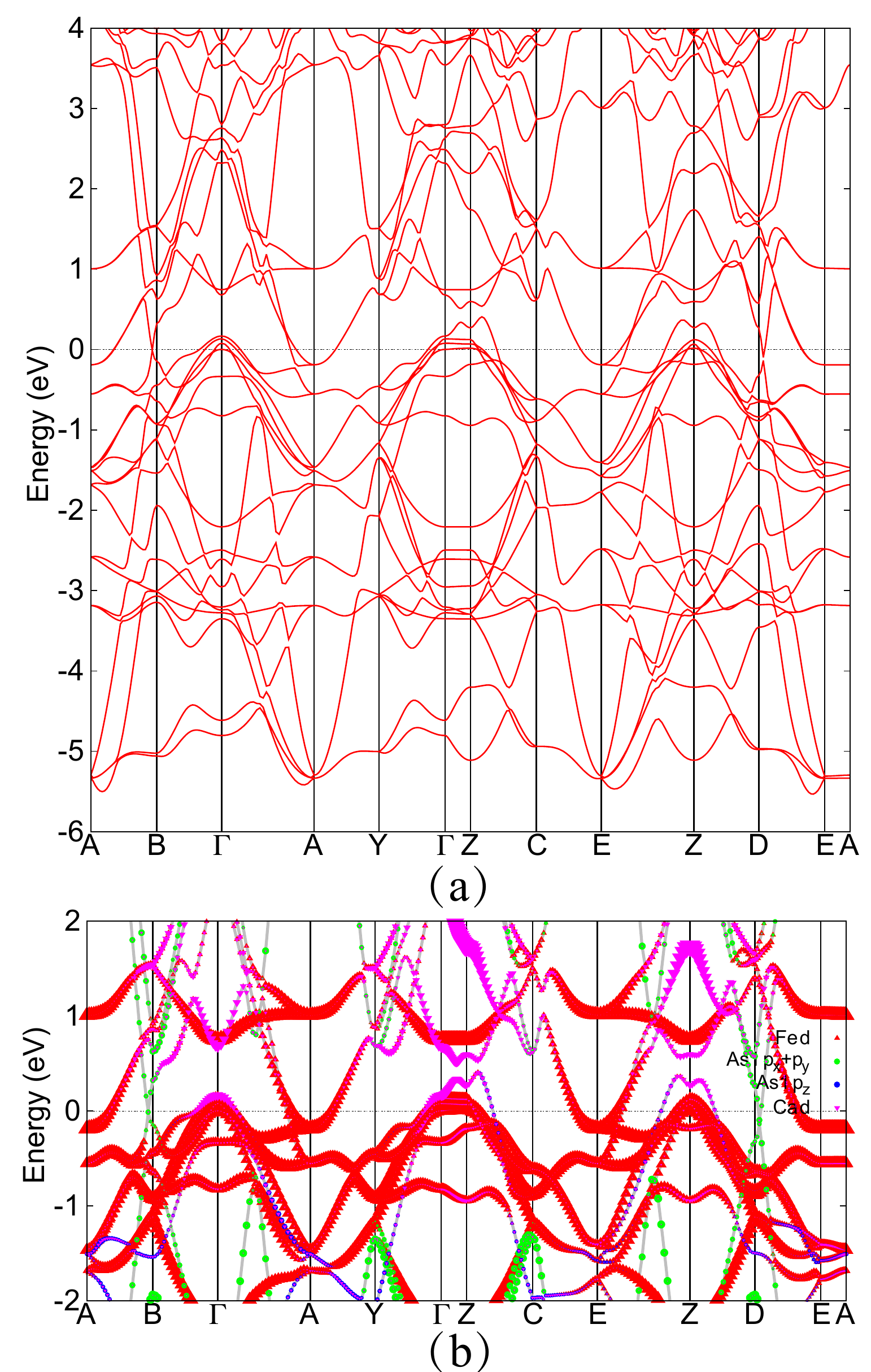}}
\caption{(color online). Band structures of CaFeAs$_2$ using GGA relaxed parameters in the paramagnetic state. The lower panel shows the orbital characters of bands near Fermi level.
 \label{band} }
\end{figure}

\begin{figure}[t]
\centerline{\includegraphics[height=4.5 cm]{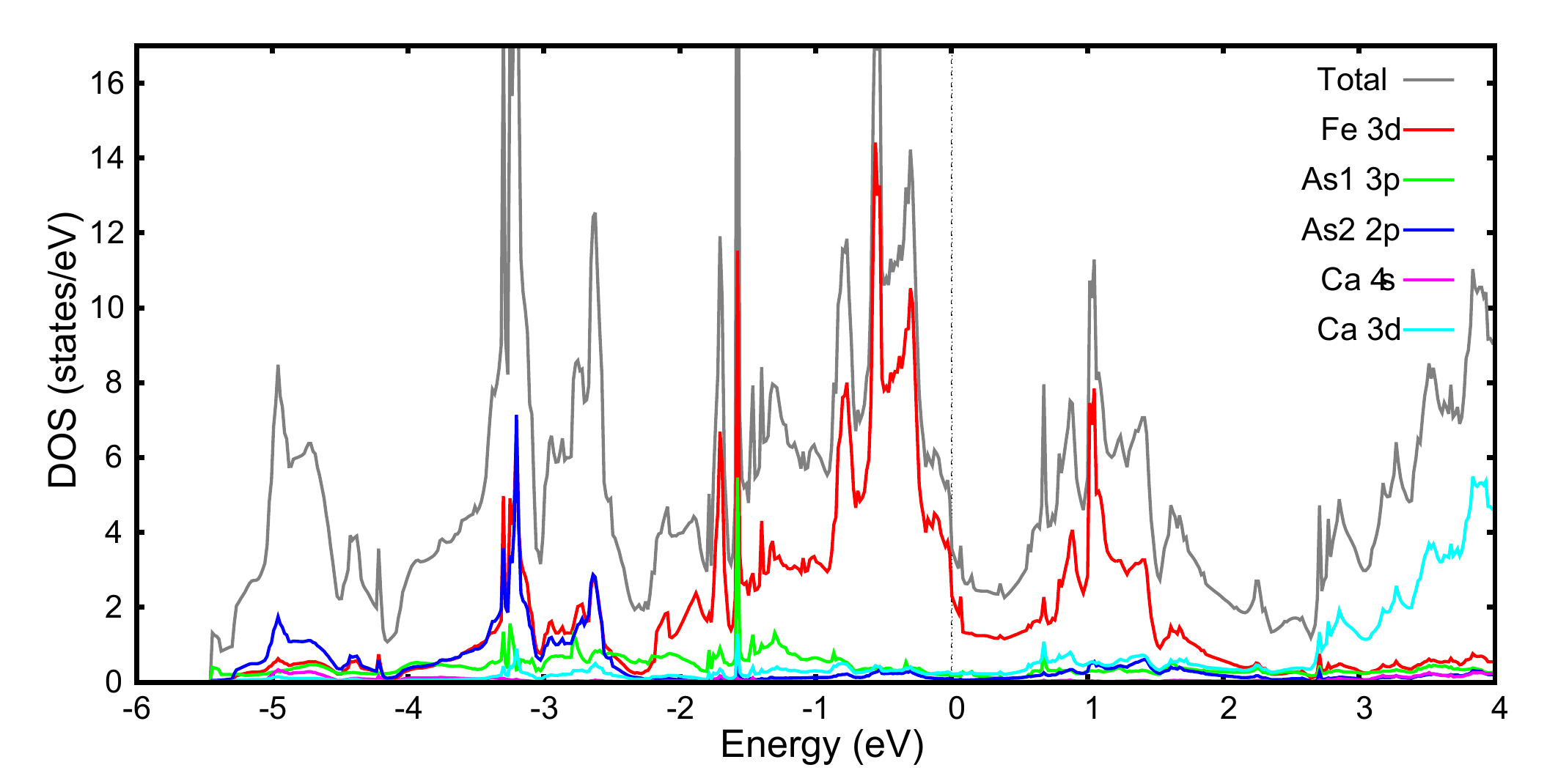}} \caption{(color online). Projected density of states for CaFeAs$_2$ in paramagnetic state using GGA.
 \label{dos} }
\end{figure}

\begin{figure}[t]
\centerline{\includegraphics[height=4.5 cm]{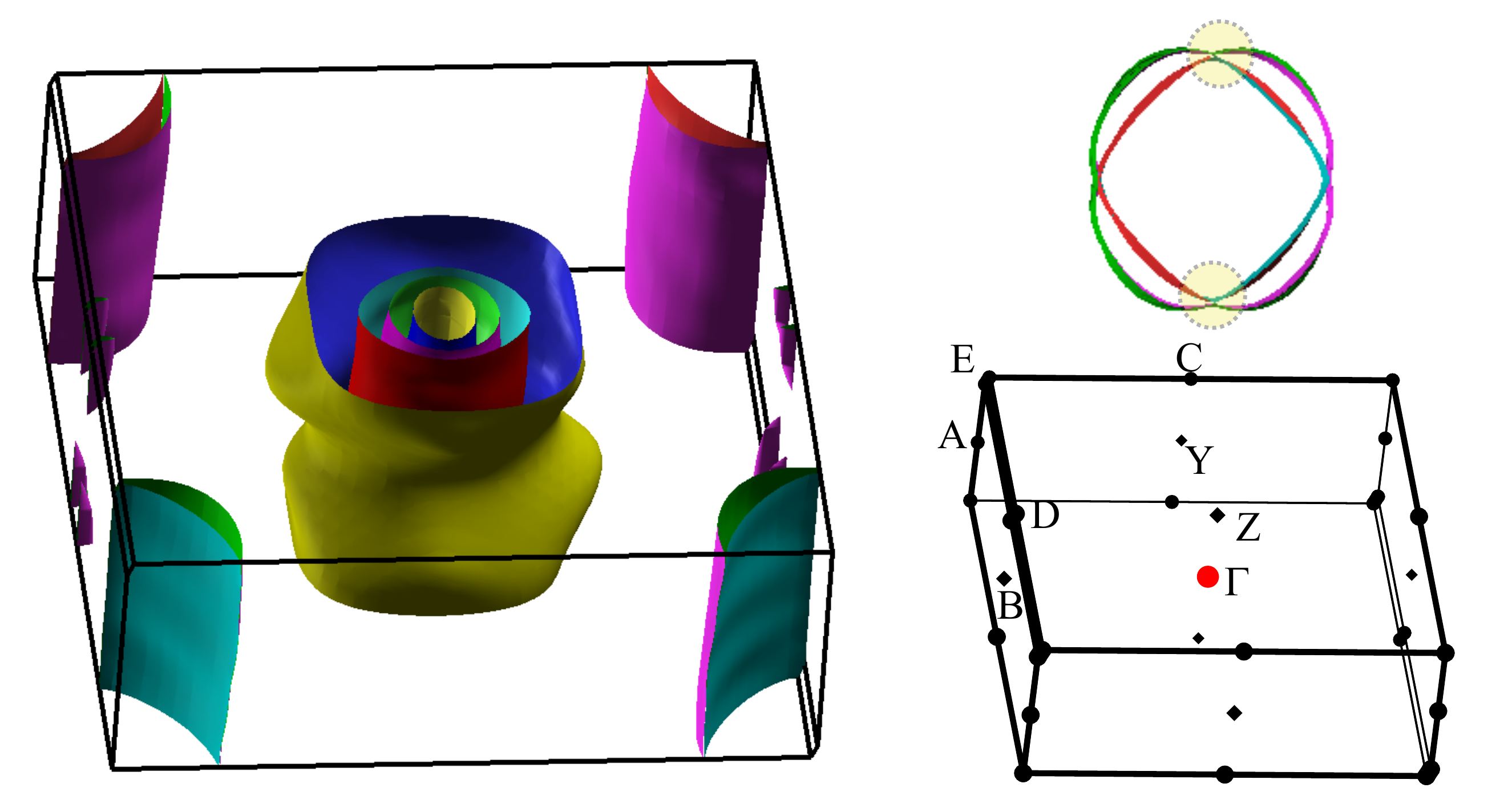}}
\caption{(color online). Calculated Fermi surface using GGA optimized lattice parameters. The right panel shows the electron Fermi surface and Brillouin zone. The shaded circles show the gap opening in $k_x=\pi$ line on electron Fermi surfaces.
 \label{fs} }
\end{figure}

The band structure and density of states(DOS) for CaFeAs$_2$ with GGA optimized structural parameters are shown in Fig.\ref{band} and Fig.\ref{dos}, respectively. Similar to iron pnictides, it is a bad metal with high density of states. The 3$d$ states of Fe are mainly located near the Fermi level from -2.5 eV to 2.0 eV and a pseudogap appears at an electron count of six. The As2 $p$ states
mainly lie 2.5 eV below the Fermi level and
are modestly coupled with the Fe 3d states. While, the As1 $p$ states extend from -4 eV to 1 eV and form dispersive bands, which indicates the $p$ orbitals of As1 are much more extended than those of As2. Despite the large separation between As1 and Fe atoms, the $p$ orbitals of As1 hybridize with 3$d$ orbitals of Fe and Ca in the energy range from -1.8 eV to 1.3 eV. While, the O $p$ states in LaOFeAs are well below the Fermi level. There is a energy shift between the $p$ states of O in LaFeAsO and As1 in CaFeAs$_2$. To analyze the orbital characters near the Fermi level, we plot the fat band, shown in Fig.\ref{band}(b). In contrast to the normal iron based material band structure, the dirac cone-like bands appear near B and D. Fig.\ref{fs} shows the calculated Fermi surfaces. There are two electron cylinders at the zone corner, four electron cones around the zone edge D-B line and three hole cylinders around zone center, and an additional 3D big hole pocket. The Fermi surfaces are very similar to the normal Fermi surfaces in iron based superconductor except for the additional 3D hole pocket and four electron cones. The additional 3D hole pocket is mainly attributed to the Ca $d$ and As1 $p_z$ states, which hybridize sufficiently with Fe $d$ orbitals. The four electron cones are mainly derived from the As1 $p_x$ and $p_y$ orbitals.

The space group of CaFeAs$_2$ consists of only a screw axis and identity, which shows clear difference compared with other Fe-based materials. The consequence of the low symmetry is that the degeneracy is preserved in $k_y=\pi$ plane but lifted in $k_x=\pi$ plane. Therefore,
the electron pockets are no longer degenerate and a small gap is opened in $k_x=\pi$ plane, shown in Fig.\ref{fs}. It may have a significant effect on the connection of electron Fermi surfaces. The band degeneracy in $k_y=\pi$ plane is protected by a hidden symmetry and the corresponding operator is,
\begin{eqnarray}
\hat{\Upsilon}=\hat{T}\hat{R}_y,
\end{eqnarray}
where $\hat{T}$ is the time reverse operator($\hat{T}$ is complex conjugate operator) and $\hat{R}_y$ is the screw operation, that is, twofold rotation with $y$ axis followed by a translation of half of the lattice vector along $y$ direction. Thus, $\hat{\Upsilon}$ is an antiunitary operator. In paramagnetic phase, $\hat{\Upsilon}H\hat{\Upsilon}^{-1}=H$ and $\textbf{k}$ in $k_y=\pi$ plane is invariant under $\hat{\Upsilon}$ operation. Furthermore, $\hat{\Upsilon}^2=-1$ at the $\textbf{k}$ in $k_y=\pi$ plane. Therefore, the bands in the $k_y=\pi$ plane are two-fold degenerate\cite{Hou2013}.

The superconductivity is likely related to the magnetism in iron
based materials. The calculated N($E_F$) with GGA optimized and experiment parameters are 3.55 and 5.08 eV$^{-1}$/f.u., respectively. These values are larger than the corresponding values of LaOFeAs (2.62
eV$^{-1}$/f.u.)\cite{Singh2008L}. N($E_F$) is large enough to put
this material near magnetism. The calculated Pauli susceptibility and specific heat coefficient with GGA optimized lattice parameters (experimental lattice parameters) are $\chi_0=$ 1.15(1.65)$\times 10^{-4} emu/mol$ and $\gamma=$ 8.8 (17.1) mJ/($K^2$ mol). To investigate the magnetic ground state of CaFeAs$_2$, we calculated the total energies of different magnetic states, which are summarized in Table.\ref{magnetic}. The collinear antiferromagnetic(AFM) state is the ground state, with an energy gain of 35.0 meV/Fe relative to a nomagnetic state and a spin moment of 1.52 $\mu_B$. We find a very weak instability of the non-spin polarized state to checkboard antiferromagnetism (1 meV/Fe, 0.79 $\mu_B$). No ferromagnetic instability is found in this material with GGA optimized parameters.

\begin{figure}
\centerline{\includegraphics[height=12 cm]{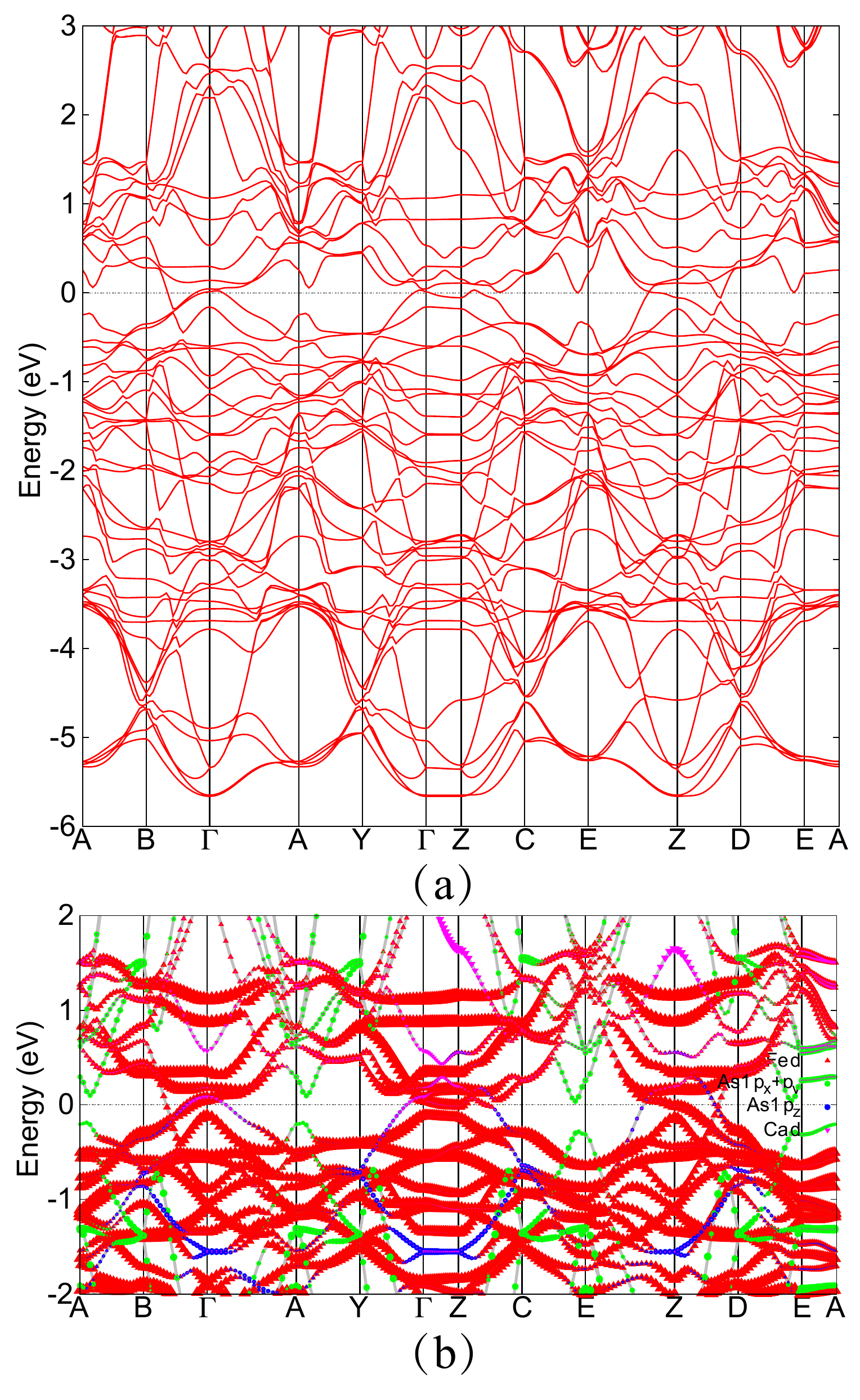}}
\caption{(color online). Band structures of CaFeAs$_2$ using GGA relaxed parameters in the collinear AFM state. The lower panel shows the orbital characters of bands near Fermi level.
 \label{band_AFM} }
\end{figure}

\begin{figure}
\centerline{\includegraphics[height=4.5 cm]{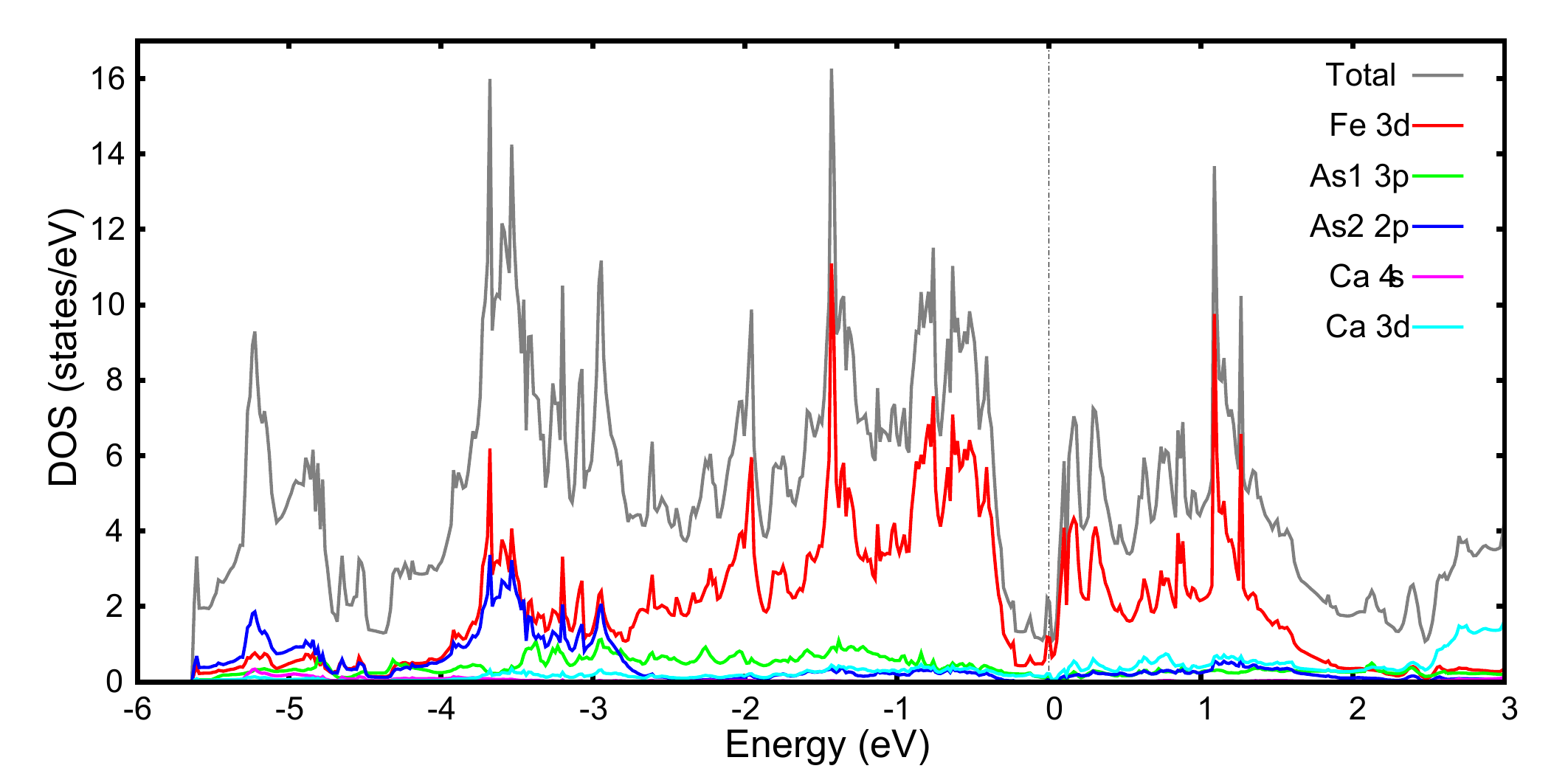}} \caption{(color online). Projected density of states for CaFeAs$_2$ in the collinear AFM state. Only the spin-up states are shown.
 \label{dos_AFM} }
\end{figure}

\begin{figure}[t]
\centerline{\includegraphics[height=4.5 cm]{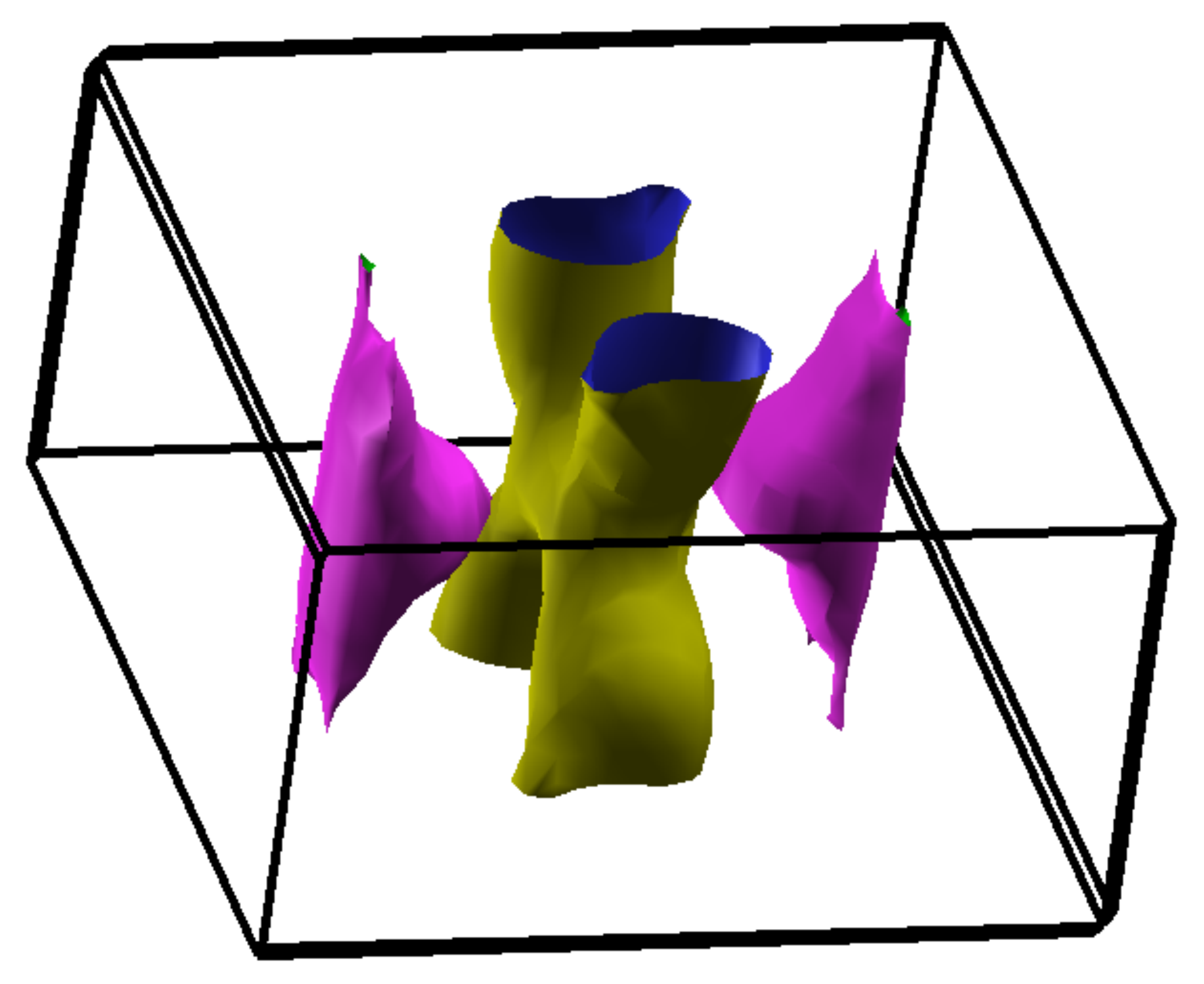}}
\caption{(color online). Calculated Fermi surface for the collinear AFM state.
 \label{fs_AFM} }
\end{figure}

Fig.\ref{band_AFM} and \ref{dos_AFM} show the calculated  band structure and DOS in the collinear AFM state (with $\sqrt{2}\times\sqrt{2}$ supercell) with internal coordinates fixed to the values obtained by non-spin-polarized energy minimization. The collinear AFM CaFeAs$_2$ is also metallic with low carrier concentration and N($E_F$) decreases severely.  Small gaps are opened at the Dirac points in the collinear AFM state. Fig.\ref{fs_AFM} shows the Fermi surfaces of CaFeAs$_2$ in the collinear AFM state. There are an "X" shape hole pocket around $\Gamma$ point and two electron pockets along $\Gamma$-B direction, which is in sharp contrast with those in  the paramagnetic state.


\begin{table}[bt]
\caption{\label{magnetic}%
The total energies for different magnetic states. They are given
relative to the total energy of the nomagnetic state and the unit is
meV/Fe.}
\begin{ruledtabular}
\begin{tabular}{cccc}
 & Nomagnetic & Checkerboard & Collinear \\
 \colrule
relative energy (meV). & 0  & -1.0 & -35.0  \\
magnetic moment ($\mu_B$). & 0  & 0.79 & 1.52  \\
\end{tabular}
\end{ruledtabular}
\end{table}


\section{tight binding model} \label{S2}
\begin{figure}[t]
\centerline{\includegraphics[height=14 cm]{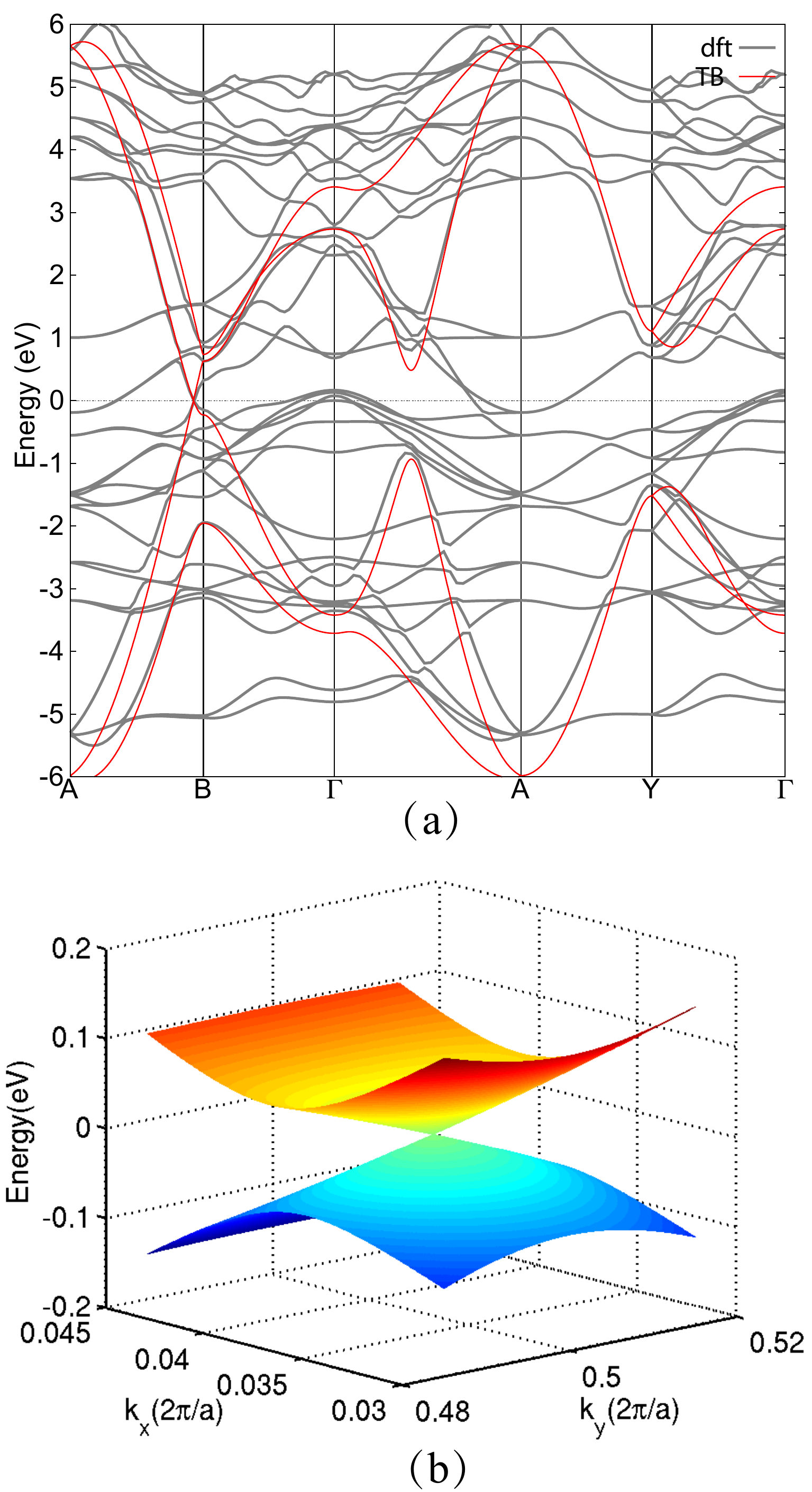}}
\caption{(color online). (a) The bandstructures of DFT and tight binding for the CaAs layers. The dark and red lines denote the DFT and tight binding bands, respectively. (b) The band dispersion near the dirac point.
 \label{dft_tb_CaAs} }
\end{figure}

From the above calculations, we see that the  four electron cones, which are attributed to the As1 $p_x$ and $p_y$ orbitals, are almost decoupled from the As1 $p_z$ orbials and the main bands of $FeAs$ layers, and  they have very small $k_z$ dispersion.   Therefore,  we  can derive a two-dimensional 4-band model  to describe the four electron cones from the CaAs layers by including the $p_x$ and $p_y$ orbitals of two As1 atoms. We introduce the operator $\phi^\dag_{\textbf{k}\sigma}=[c^\dag_{ax\sigma}(\textbf{k}),c^\dag_{ay\sigma}(\textbf{k}),c^\dag_{bx\sigma}(\textbf{k}),c^\dag_{by\sigma}(\textbf{k})]$, where $c^\dag_{\alpha \beta\sigma}$ creates a $p_\beta$($\beta=x,y$) electron with spin $\sigma$, sublattice $\alpha$($\alpha=a,b$) and wave vector $\textbf{k}$. Then, the tight-binding Hamiltonian can be written as: $H=\sum_{\textbf{k}\sigma}\phi^\dag_{\textbf{k}\sigma}h(\textbf{k})\phi_{\textbf{k}\sigma}$. The Hamiltonian matrix is,
\begin{eqnarray}
\label{caas_tb}
&&h_{11}=h_{33}=\epsilon_X+2t^{11}_{1}cosk_x +2t^{11}_{2}cosk_y, \nonumber\\
&&h_{13}=(2t^{13}_{1}e^{i(x_0-1)k_x}+2t^{13}_{2}e^{ix_0k_x})cos(k_y/2),\nonumber\\
&&h_{14}=h_{23}=-(2it^{14}_1e^{i(x_0-1)k_x}+2it^{14}_2e^{ix_0k_x})sin(k_y/2),\nonumber\\
&&h_{22}=h_{44}=\epsilon_Y+2t^{11}_{2}cosk_x +2t^{11}_{1}cosk_y,\nonumber\\
&&h_{24}=(2t^{24}_1e^{i(x_0-1)k_x}+2t^{24}_2e^{ix_0k_x})cos(k_y/2),
\end{eqnarray}
where $x_0=0.42$ is the difference between the $x$ components of $a$ and $b$ sublattices. The corresponding tight binding parameters are (all in eV),
\begin{eqnarray}
&&\epsilon_X=-0.30, \quad \epsilon_Y=-0.109, \quad t^{11}_1=-0.149,  \nonumber \\
&&t^{11}_2=0.128, \quad t^{13}_1=0.89, \quad t^{13}_2=0.649, \nonumber \\
&&t^{14}_1=1.169,\quad t^{14}_2=-1.740, \quad t^{24}_1=0.567  \nonumber \\
&& t^{24}_2=1.213.
\end{eqnarray}
The DFT and tight binding bands are shown in Fig.\ref{dft_tb_CaAs}(a) and they match well near Fermi level. From Eq.\ref{caas_tb}, we find that the intraorbital couplings vanish for $k_y=\pi$ and each band in $k_y=\pi$ plane is two-fold degenerate. While, the inter and intra orbital couplings between $a$ and $b$ sublattices do not vanish for $k_x=\pi$ and the bands in $k_x=\pi$ are not degenerate. They differ from the bands in iron based superconductors with $P4/nmm$ and $I4/mmm$ space groups, where all bands in both $k_x=\pi$ and $k_y=\pi$ planes are two-fold degenerate. Consider the k point $k=(\pi,k_y)$$(k_y\rightarrow0)$ in $A$-$B$ line, the eigenvalues of the Hamiltonian $E\sim\pm2(t^{14}_1-t^{14}_2)k_y$(only consider dominant terms). Therefore, the two linearly dispersing bands cross each other around $B$, forming a Dirac cone. It is consistent with DFT calculation. The anisotropic Dirac cone obtained by tight binding model is shown in Fig.\ref{dft_tb_CaAs}(b).

Similar to other iron-based superconductors,  the electronic structure of the FeAs layers  has contribution from all d-orbitals at iron atoms. However, as the As1 $p_z$ contributes an additional hole pocket with large $k_z$ dispersion,  the coupling between the Fe d orbitals and  the As1 $p_z$ can not be ignored.  Considering the symmetry breaking of FeAs layers is not too severe, we can construct a six-band tight binding model including five $d$ orbitals of Fe and $p_z$ orbital of As1 to describe the main band structures. The hopping parameters are listed in Table.\ref{hopping}. The onsite energies $\epsilon_i$ (in eV) $\epsilon_1=-0.0411$, $\epsilon_3=-0.329$, $\epsilon_4=0.122$, $\epsilon_5=-0.182$, and $\epsilon_6=-1.591$ where $i$=1 corresponds to the $d_{xz}$, $i$=2 to the $d_{yz}$, $i$=3 to the $d_{x^2-y^2}$, $i$=4 to the $d_{xy}$, $i$=5 to the $d_{3z^2-r^2}$, and $i$=6 to the As1 $p_z$.

\begin{figure}[t]
\centerline{\includegraphics[height=8 cm]{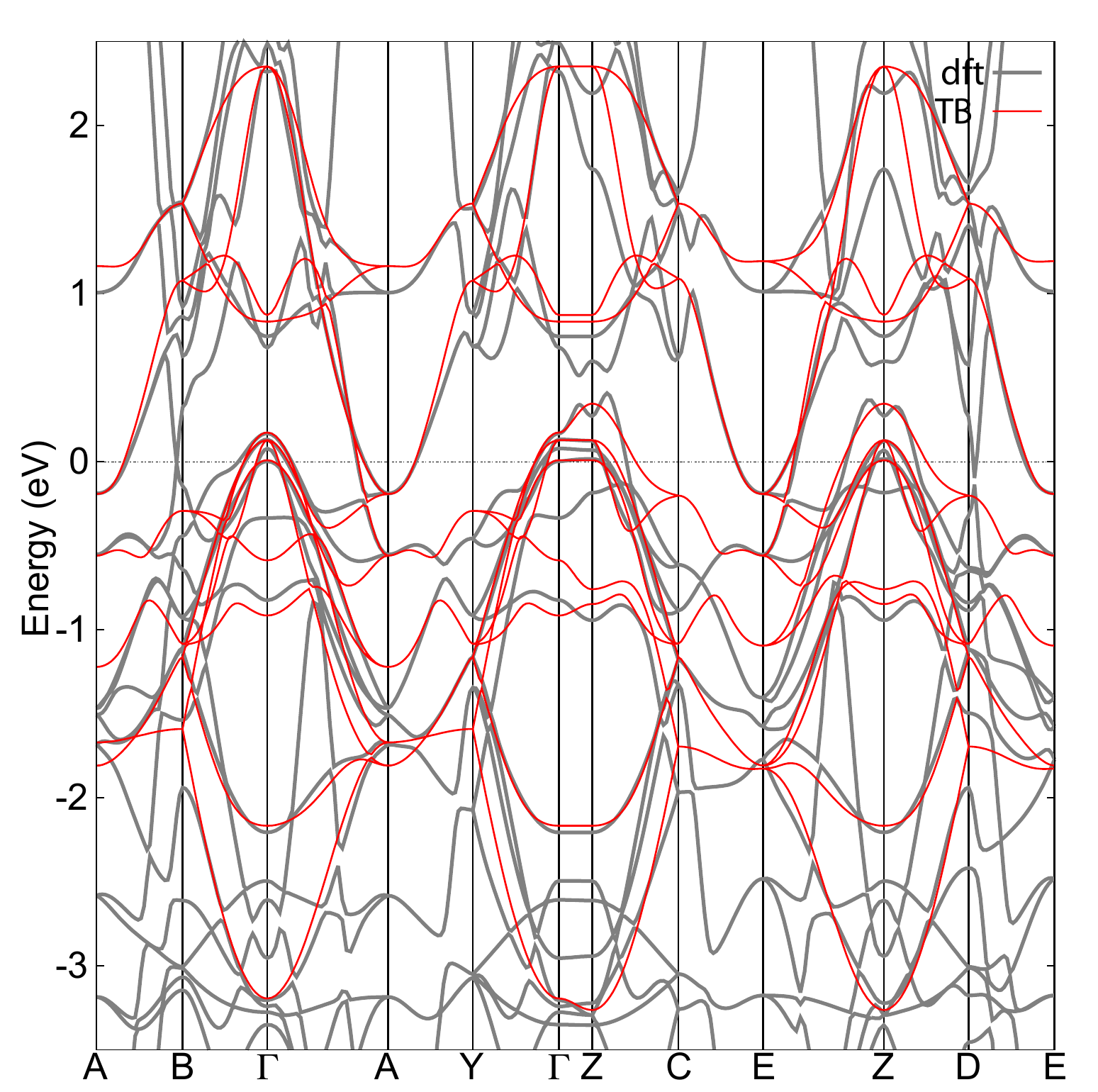}}
\caption{(color online). The bandstructures of DFT and tight binding. The dark and red lines denote the DFT and tight binding bands, respectively.
 \label{dft_tb} }
\end{figure}

\begin{table}[bt]
\caption{\label{hopping} The hopping parameters  to fit  the DFT results in  the six-orbital model. The definition of the kinetic energy terms $\xi_{mn}(k)$ follow those in Ref.\onlinecite{Graser2009}.
 The $z$ in parentheses denotes the hopping $t^{mn}_z$. The $x$ direction is along the Fe-Fe bond.}
\begin{ruledtabular}
\begin{tabular}{ccccccccc}
$t^{mn}_i$ & $i$=x & $i$=y & $i$=xy & $i$=xx & $i$=yy & $i$=xxy & $i$=xyy & $i$=xxyy \\
 \colrule
$mn$=11 & -0.08  & -0.40 & 0.28 & 0.02 & -0.01 & -0.040 &  & 0.035  \\
$mn$=33 & 0.375  &   +$t_x$   & -0.075 & -0.022 &  +$t_{xx}$ &   &    & 0.013  \\
$mn$=44 & 0.172  &   +$t_x$    & 0.125 & -0.025 &  +$t_{xx}$  & -0.032  & +$t_{xyy}$  & -0.02  \\
$mn$=55 & -0.061  &   +$t_x$    & -0.095 & -0.042 & +$t_{xx}$  & 0.01  & +$t_{xxy}$  & -0.006  \\
$mn$=66 & -0.421  &   +$t_x$    & 0.02 &   &    &    &    &    \\
$mn$=12 &     &       & 0.121 &   &    & -0.022  &  +$t_{xxy}$ & 0.043  \\
$mn$=13 &    & -0.414  & 0.112 &   &    & 0.022  & -$t_{xxy}$  &   \\
$mn$=14 & -0.32  &       & -0.015 & -0.007 &   & -0.018  &   &   \\
$mn$=15 & -0.093  &       & -0.103 &   &    &    &    & -0.021  \\
$mn$=34 &    &       &   &   &    & 0.01  & -$t_{xxy}$  &    \\
$mn$=35 & -0.334  &  -$t_x$    &    &    &     & -0.018  &  -$t_{xxy}$ &   \\
$mn$=45 &    &       & 0.121 &    &     &     &    & -0.012  \\
$mn$=56 & 0.2(z)  &      &  &  &   &    &   &    \\
\end{tabular}
\end{ruledtabular}

\end{table}


\begin{eqnarray}
\xi_{66}&=&2t^{66}_x(cosk_x+cosk_y)+4t^{66}_{xy}cosk_x cosk_y,\nonumber\\
\xi_{56}&=&2it^{56}_zsin(k_z/2).
\end{eqnarray}
The tight binding band fits the DFT band well near the Fermi level, shown in  Fig.\ref{dft_tb}. Because of the symmetry breaking, the bands in A-B and D-E lines are no longer degenerate. This effect can be modeled by adding a symmetry breaking term $sink_x+sink_y$ in $\xi_{12}$.

\section{Discussion} \label{S3}
Unlike LaFeAsO, the spacer layers in CaFeAs$_2$ can contribute to the states near Fermi level and are metallic. The As1 $p_z$ orbitals can couple with Fe $d$ orbitals, which make this material less two dimensional than LaOFeAs.  These unique feature may be used to test the role of charge and polarization fluctuation, the importance of two dimensionality and pairing symmetry in Fe-based superconductors\cite{Shim2009}.

\begin{figure}[t]
\centerline{\includegraphics[height=8 cm]{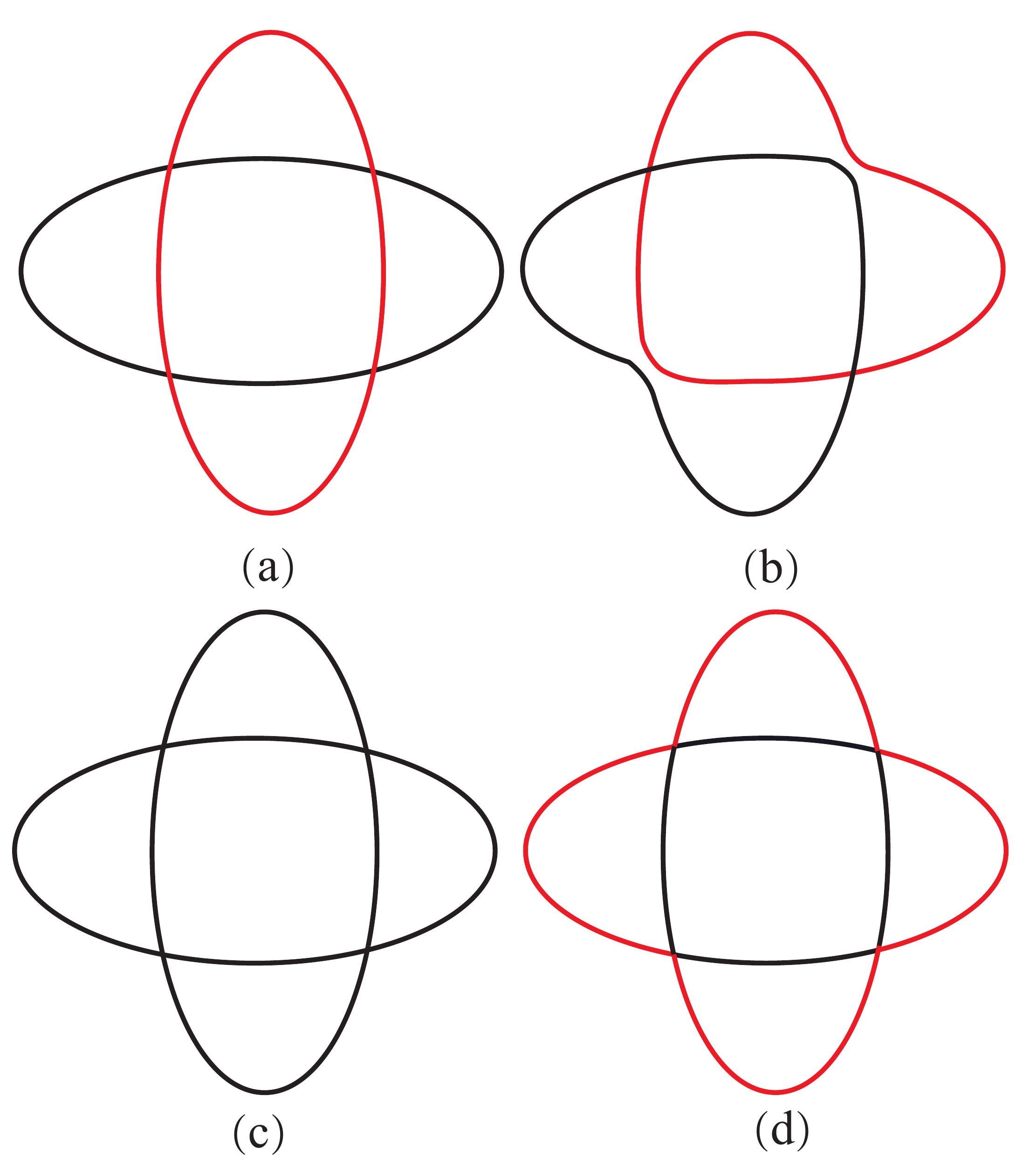}}
\caption{(color online). (a) The band connection of electron Fermi surfaces in normal iron based materials; (b) the band connection of electron Fermi surfaces in CaFeAs$_2$. The curves with different colors denote different bands. (c) The phase of normal SC state on electron Fermi surfaces. (d) The phase of anti-phase SC state  on electron Fermi surfaces. The sign difference is indicated by black and red colors.
 \label{fs_con} }
\end{figure}

In previous discovered iron based materials, the electron Fermi surface consists of two ellipses and the band connection is shown in Fig.\ref{fs_con}(a). In CaFeAs$_2$ system, however, the degeneracy is lifted in  the $k_x=\pi$ plane and the band connection should have a significant difference from the previous discovered materials.  The topology of the electron pockets in CaFeAs$_2$ appears as   Fig.\ref{fs_con} (b). From our calculation, the separation between the two pockets near $k_F$ is about 7 meV and  slightly increases with hole doping.

The pairing symmetry in iron-based superconductors has not been settled\cite{hirschfeld}. There are many different proposals.
The degeneracy lift  is manifested by the coupling between the two electron bands.  The coupling can help to determine some features related to pairing symmetries. One essential debate about pairing symmetries is whether there is a sign change among electron pockets. For example, the sign change exists in both the d-wave\cite{maier} and the anti-phase s$^\pm$ pairing
symmetry\cite{Hus4, Hu2013,Hao2013,mazin2,khodas} while  it is absent in a standard s-wave or s$^\pm$\cite{mazin,kuroki,seo, wang}. If such a sign change exists, the small coupling can cause very interesting effect on the gap functions on  the electron pockets.  To see this, one can write a general  meanfield Hamiltonian for the two electron bands as
\begin{eqnarray}
&&H=\sum_{\textbf{k}}\Psi^{\dag}(\textbf{k})h(\textbf{k})\Psi(\textbf{k}),\nonumber\\
&&h(\textbf{k})=\left(
\begin{array}{cccc}
\epsilon_1(\textbf{k}) & \Delta_1(\textbf{k}) & V(\textbf{k}) & 0  \\
\Delta_1(\textbf{k}) & -\epsilon_1(\textbf{k}) & 0 & -V(\textbf{k}) \\
 V(\textbf{k})  &  0 & \epsilon_2(\textbf{k}) & \Delta_2(\textbf{k}) \\
   0   &  -V(\textbf{k}) &  \Delta_2(\textbf{k}) & -\epsilon_2(\textbf{k})  \\
\end{array}
\right),
\end{eqnarray}
with $\Psi^{\dag}(\textbf{k})=(c^{\dag}_{1\textbf{k}\uparrow},c_{1-\textbf{k}\downarrow},c^{\dag}_{2\textbf{k}\uparrow},c_{2-\textbf{k}\downarrow})$ in Nambu spinor representation. $c^\dag_{\alpha\textbf{k}\sigma}$ creates an electron with wave vector $\textbf{k}$, band $\alpha$ and spin $\sigma$. $V(\textbf{k})$ is the inter band coupling due to the low crystal symmetry of CaFeAs$_2$. Near the Fermi level, $\epsilon_{1,2}(\textbf{k})=0$ and the four Bogoliubov quasi-particle eigenvalues $E_1 = -E_3$ and $E_2 = -E_4$, which are given by
\begin{eqnarray}
E_{m=1,2}(\textbf{k})=\frac{1}{2}[\pm(\Delta_1-\Delta_2)+\sqrt{(\Delta_1+\Delta_2)^2+4V^2}].
\end{eqnarray}
For simplicity, we take $|\Delta_1|=|\Delta_2|$, which will not affect our qualitative analysis. If the two order parameters have the same sign,  $E_{m=1,2}(\textbf{k})=\sqrt{\Delta_1^2+V^2}$. The coupling has little effect on the gap functions.  However, if the sign change exists  between two electron pockets\cite{Hus4, Hu2013,Hao2013}, $E_{m=1,2}(\textbf{k})=\pm\Delta_1+V$.  In this case,  if the original gap functions on the electron Fermi surfaces are close to isotropic as measured in many other iron-based superconductors, the coupling is expected to  create large anisotropic gaps among electron pockets. Therefore, if  there is a sign change between the electron pockets, a large anisotropic gap can be expected on electron pockets in CaFeAs$_2$.

The existence of the additional hole pocket also may provide a way to  determine the  s$^\pm$ pairing symmetry. The additional hole pocket has large c-axis dispersion. It is expected to dominate the c-axis transportation. Therefore, on the CaAs layer, the superconducting order should have  the same sign as the hole pockets of the  FeAs layers. This feature allows to construct a  $\pi$-junction device  to access the sign of electron pockets and hole pockets separately as suggested in Ref.\onlinecite{tsai}.

\section{Conclusion} \label{S4}
In summary, we report the electronic structures and magnetic properties for  CaFeAs$_2$. The recently discovered Iron-based superconductors have chain-like As layers. We find that the states of CaAs blocking layers, compared with LaO layers in LaOFeAs, are metallic and couple strongly with those of the FeAs layers. The Fermi surfaces are similar to those of LaOFeAs except for the additional 3D hole pocket and four electron cones. The additional hole pocket is mainly attributed to the Ca $d$ and As1 $p_z$ states and the electron cones are derived from the As1 $p_x$ and $p_y$ states. The ground state of CaFeAs$_2$ is found to be a collinear antiferromagnetic semimetal. Ignoring the electron cones, this materials can be well described  by a six-band model, including five Fe $d$ and As1 $p_z$ orbitals. Moreover due to the low symmetry crystal induced by the  As layers, the bands attributed to FeAs layers in $k_y=\pi$ plane are two-fold degenerate but in $k_x=\pi$ plane are lifted. This degeneracy is protected by an hidden symmetry $\hat{\Upsilon}=\hat{T}\hat{R}_y$. It causes the band rearrangement in $k_x=\pi$ plane and the band connection of electron Fermi surface has changed. This feature and the additional hole pocket may be used to identify the sign change and pairing symmetry of iron based superconductors and shed light on the mechanism underlying high-temperature superconductivity .

\section{Acknowledgments}

The work is supported by "973" program (Grant No.
 2010CB922904 and No. 2012CV821400), as well as  national science foundation of China (Grant No. NSFC-1190024, 11175248 and 11104339).

\end{document}